# Title: Unsupervised image registration towards enhancing performance and explainability in cardiac and brain image analysis

Chengjia Wang [1], Guang Yang* [2], Giorgos Papanastasiou* [1, 3]

[1] Edinburgh Imaging Facility QMRI, Centre for Cardiovascular Science, University of Edinburgh, Edinburgh, UK

[2] Faculty of Medicine, National Heart & Lung Institute, Imperial College London, London, UK

[3] School of Computer Science and Electronic Engineering, University of Essex, UK

E-mail: chengjia.wang@hw.ac.uk, g.yang@imperial.ac.uk, g.papanastasiou@essex.ac.uk

* Co-corresponding authors

**Abstract**

Magnetic Resonance Imaging (MRI) typically recruits multiple sequences (defined here as "modalities"). As each modality is designed to offer different anatomical and functional clinical information, there are evident disparities in the imaging content across modalities. Inter- and intra -modality affine and non-rigid image registration is an essential medical image analysis process in clinical imaging, as for example before imaging biomarkers need to be derived and clinically evaluated across different MRI modalities, time phases and slices. Although commomly needed in real clinical scenarios, affine and non-rigid image registration is not extensively investigated using a single unsupervised model architecture. In our work, we present an unsupervised deep learning registration methodology which can accurately model affine and non-rigid transformations, simultaneously. Moreover, inverse-consistency is a fundamental inter-modality registration property that is not considered in deep learning registration algorithms. To address inverse-consistency, our methodology performs bi-directional cross-modality image synthesis to learn modality-invariant latent representations, while involves two factorised transformation networks (one per each encoder-decoder channel) and an inverse-consistency loss to learn topology-preserving anatomical transformations. Overall, our model (named "FIRE") shows improved performances against the reference standard baseline method (i.e. Symmetric Normalization implemented using the ANTs toolbox) on multi-modality brain 2D and 3D MRI and intra-modality cardiac 4D MRI data experiments. We focus on explaining model-data components to enhance model explainability in medical image registration. On computational time experiments, we show that the FIRE model performs on a memory-saving mode, as it can inherently learn topology-preserving image registration directly in the training phase. We therefore demonstrate an efficient and versatile registration technique that can have merit in multi-modal image registrations, in the clinical setting.

**Keywords**: Multi-modality image registration; Unsupervised image registration; Deep learning; Inverse-consistency; Explainable deep learning

## 1. Introduction

Clinical decision making from magnetic resonance imaging (MRI) is based on combining anatomical and functional information across multiple MRI sequences (defined throughout as "modalities"). Multiple imaging biomarkers can be derived across different MR modalities and organ areas. This makes image registration an important MR image analysis process, as it is commonly required to "pair" images from different modalities, time points and slices. Hence, intra- as well as inter-modality image registration are essential components in clinical MR image analysis [1], and finds wide use in longitudinal analysis and multi-modal image fusion [2].

Although numerous deep learning (DL) methods have been devised for medical image analysis [2], DL-based image registration tasks have been relatively less explored [3, 4]. Among DL-based registration studies, supervised learning-based methods showed promising results. The main disadvantage of supervised-learning is that it necessitates laborious and time-consuming manual annotations for model training, whilst it is difficult to design generalized frameworks [5-7]. Unsupervised DL-based image registration methods gain increasing popularity, as they aim to overcome the need of training datasets with ground truth annotations [3, 4]. However, unsupervised learning has been mainly investigated on intra(single)-modal image registration [3, 4, 8] and on either 3D volumes [9-13] or 2D images [14-18]. Also, previous unsupervised learning methods involve affine registration before training, which is a laborious, time-consuming and computationally expensive task [2-4, 9-13, 14-18].

To date, most previous unsupervised DL methods can perform either affine or non-rigid registration [3, 4, 9, 10, 12, 13, 14-18]. To our knowledge, the only unsupervised DL study that has developed a method for both affine and non-rigid registrations is by de Vos et al [11]. In this work, DL modelling was performed using two autonomous models in the analysis pipeline to address both affine and non-rigid registrations, whilst also required affine transformations before training. Inverse-consistency-based transformation models show improved ability to preserve the contextual and topology information in image registration [19]. Note that a deformable image registration between two images is called inverse consistent, when the correspondence between images is invariant to the order of the choice of the source and target image [19-21].

In our work, we demonstrate a bi-directional unsupervised DL model that is capable to perform multi-modal (n-D, where n=2-4) affine and non-rigid image transformations. As it will be discussed in the next sections, we model n-D affine and non-rigid registrations through bi-directional cross-modality synthesis and inverse-consistent spatial transformations. Unlike previous unsupervised learning studies that have focused on estimating asymmetric transformations and cannot preserve topology [2-4, 9-13, 14-18], our proposed technique is efficient for both affine and non-rigid image registrations, as demonstrated on multi-modal brain and cardiac image registration experiments.

**1.1 Motivation**

Unsupervised DL techniques have considerable potential in medical image registration. In this paper we combine up-to-date progress in unsupervised learning with strong prior knowledge about medical image registration: reaching high registration performance, we synchronically perform multi-modal image synthesis and factorised spatial transformations (one per each encoder-decoder channel). This operation allows to efficiently reach inverse-consistent multi-modal affine and non-rigid image registration of multi-dimensional medical imaging data, making our work efficient and versatile for the medical imaging community.

The FIRE model provides a generalised architecture for registration applications, as the synthesis components of the FIRE model can be customized and re-trained, if/when data from additional medical imaging modalities would need to be co-registered. The synthesis factor captures global image modality information, determining how organ topology is rendered in the target image. Between other things, maintaining a representation of the modality characteristics in the synthesis factor of the model, provides the ability to potentially model data from multiple modalities. We evaluate this function using multi-modal MR data of variable anatomical information (see also "Model explainability" section).

Finally, the incorporation of factorised spatial transformations in the FIRE model, allows to learn varied image transformations. We examine whether the FIRE architecture and its associated loss functions can directly output transformation fields for both affine and non-rigid registrations. We demonstrate that the FIRE model architecture efficiently learns to produce inverse-consistent deformations that are are intrinsically topology-preserving.

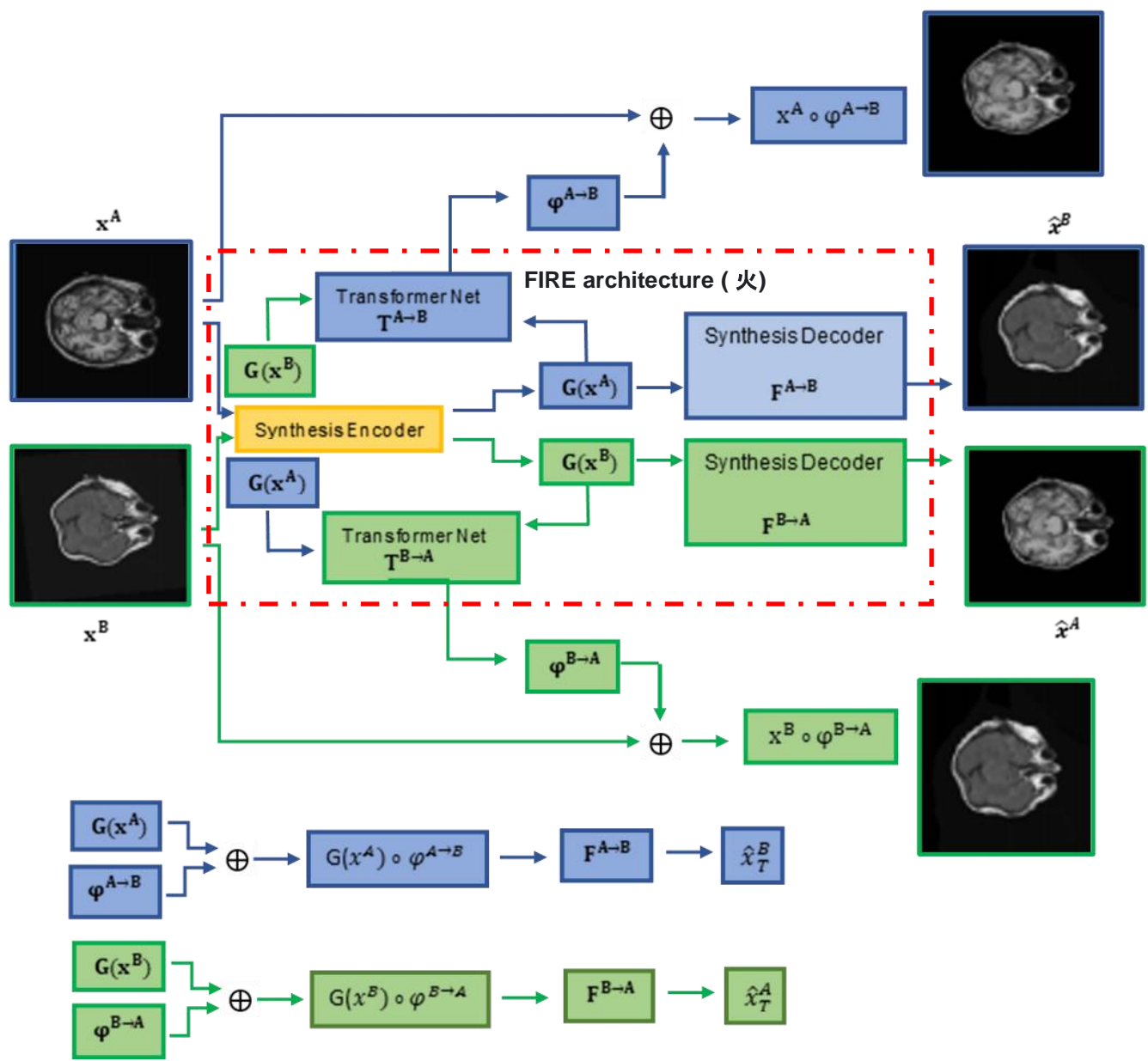

**Figure 1)** FIRE model architecture. Two synthesis encoders G ($x^A$) and G ($x^B$) extract modality-invariant (latent) representations. Two synthesis decoders $F^{A \to B}$ and $F^{B \to A}$ map the representations extracted by G ($x^A$) and G ($x^B$) to $\hat{x}^B$ and $\hat{x}^A$ (images synthesised), respectively (see $F^{A \to B}$ and $F^{B \to A}$ at the bottom). Finally, two transformation networks $T^{A \to B}$ and $T^{B \to A}$ model the transformation fields, finalising the FIRE training process.

### 1.2 Overview of the proposed approach

Learning image representations and spatial transformations through image synthesis, is an area of recent work in medical image analysis [20]. However, there is no previous consideration about the precision of the synthesis process and the versatility of the spatial transformations. This is important in medical image analysis, as both affine and deformable registrations are commonly required for a typical clinical dataset.

Following our previous work [22], we demonstrate a versatile registration method called "FIRE", by explicitly incorporating spatial transformation into our cross-domain bi-directional image synthesis (Figure 1). We show that our method can robustly model 2D/ 3D and 4D registrations when examined on multi- and intra-modality brain and cardiac MRI, respectively. We provide thorough explanations regarding our model explainability in medical image registration. Moreover, we demonstrate the efficiency of our method on computational time experiments.

### 1.3 Contributions

Our main contributions are as follows:

- With the use of two spatial transformation networks factorised into our cross-domain synthesis, the "火"-shape FIRE model architecture can be trained (through our combined loss functions) end-to-end, to perform inverse-consistent unsupervised registrations, using a single DL network for both inter- and intra-modality registration.

- Incorporating a new affine regularisation loss to eliminate non-rigid transformation where needed, we show that our FIRE model learns simultaneously both affine and non-rigid transformations.

- Overall, the novel FIRE model loss functions and training procedure are designed to allow simultaneous learning for synthesis and registration of 2D, 3D and 4D multi-modal MR data.

- We show that our FIRE model is explainable and provide a comprehensive (affine and non-rigid) framework for future improvements and generalisation.

## 2. Related work

In this section we review previous work on registration methods using DL (section 2.1). We then review inverse-consistent DL transformations (section 2.2). Finally, we review methods that solve inter-modality registration trough image synthesis (section 2.3).

Since we detail the background and motivation of our work against previous studies in the Introduction section of this paper, here we focus on discussing similarities and differences of our FIRE model against previous methods that are relevant to our application domain.

### 2.1 Registration methods using DL

Early DL registration methods were mostly adaptations of the conventional image alignment scheme, where DL was used to extract convolutional features for guiding correspondence detection across subjects [23], or to learn new similarity metrics across images [24, 25].

Reinforcement learning methods have shown promising results, but they are still designed to solve image registration through iterative processes, making them computationally expensive and time-consuming [26-28]. As discussed, although DL supervised learning-based methods have also being promising for image registration, one of their main disadvantage is that they require laborious and time-consuming manual annotations [5-7, 29].

To overcome the limitations of supervised learning methods, unsupervised learning approaches are recently developed, which learn the image registration by focusing to minimise the loss between the deformed image and fixed (target) image. The development of a spatial transformer network (STN) in 2015 by Jaderberg et al. has inspired many unsupervised learning methods as it can be inserted into existing DL models [8]. The STN allows explicit manipulation of images within a network and can potentially perform image similarity loss calculations during the training and testing process. Balakrishnan et al. demonstrated Voxelmorph, a 3D medical image registration algorithm using STN within a convolutional neural network (CNN), in which parameters are learned by the local cross-correlation loss function [30, 31]. Krebs et al. proposed a probabilistic formulation of the registration problem through unsupervised learning, in which an STN was used within a conditional variational autoencoder [31]. Kim et al. developed CycleMorph, a cycle consistent DL method for deformable image registration (through an STN component) and demonstrated accurate registration results [33]. Unlike previous techniques [30-32, 34, 35], the CycleMorph model was incorporating inverse consistency directly in the training phase [33], to improve model efficiency and robustness.

To date, the only unsupervised DL study that has developed a method for both affine and non-rigid registrations is by de Vos et al [11]. However, in this work, DL modelling required the incorporation of two autonomous models in the analysis to address both affine and non-rigid registrations, whilst also required affine transformations as part of the modelling process (therefore adding complexity and computational cost).

**2.2 Inverse-consistent DL transformations**

Most recent studies are mainly focusing on performing non-rigid deformations [30-36], whilst require computationally expensive and laborious affine transformations in the pre-processing using ANTs [30, 36, 37] or some other software application [20]. Note that affine transformation is inherently inverse-consistent due to the invertibility of the transformation matrix. In the past decade, inverse-consistency has mainly been achieved by incorporating diffeomorphic deformation fields [37]. Some of the most popular examples include the diffeomorphic Demons algorithms [38, 39].

Several DL diffeomorphic architectures have recently been proposed, such as the Voxelmorph [30, 31, 34, 35], the Quicksilver [38] and the CycleMorph [33]. However, most of these unsupervised diffeomorphic models were tested on intra-modality brain data [30, 31, 34, 35, 38]. As already discussed in subsection 2.1, the CycleMorph model is still considered the only technique to effectively incorporate inverse consistency directly in the training phase, whilst it was tested on multi-modal (Brain MRI and liver CT data) medical imaging data. In our work, we define the output deformation fields with simple inverse-consistency constraints, devising a model that can learn topology-preserving registratrion in an expandable and versatile architecture. It may be important to emphasise that Nielsen et al. recently concluded that the diffeomorphism fails to account for local discontinuities in brain images with pathological conditions [39]. Despite this finding, we should highlight that it is straightforward to integrate diffeomorphic layer components into the FIRE framework if deemed as appropriate (for example, the Voxelmorph velocity field [30, 31], or the CycleMorph homeomorphic mapping [33] could potentially be incorporated).

In clinical practice, image registration tasks commonly involve comprehensive image deformations, i.e. a combination of affine and non-rigid registration. In our study, we demonstrate a versatile and efficient DL model that can perform both affine and non-rigid (deformable) registration which can learn to produce inverse-consistent and intrinsically topology-preserving image deformations.

**2.3 Image synthesis DL methods**

Image synthesis-based methods have potential in recent DL-based inter-modality registration algorithms. Alongside the recent developments on adversarial cycle-consistent loss [41], most recent methods focus on using CycleGAN for synthesis-based inter-modality registration [20, 41]. Wei et al. performed image registration by devising CycleGAN-based image synthesis as a separate processing step prior to registration [41]. Qin et al. proposed a bi-directional unsupervised training framework, which is conceptually close to our model. However, our FIRE model is simultaneously learning synthesis and registration on a memory-saving architecture that requires less hyperparameters when balancing all different losses (see Methods). Moreover, the FIRE model is evaluated on 2D, 3D and 4D data involving different organs (brain and heart; note that 4D here denotes periodic heart deformations observed across different time phases of cardiac cine-MRI).

## 3. Methods

Our proposed model can be described as a bi-directional cross-domain image synthesis structure, with two factorised spatial transformation components (one per each encoder-decoder channel) [22].

Each time two images $x^A$ and $x^B$ are processed, the FIRE model learns $\varphi^{A \to B}$ and $\varphi^{B \to A}$ transformations, warping $x^A$ and $x^B$ into $x^A \circ \varphi^{A \to B}$ and $x^B \circ \varphi^{B \to A}$, via the transformation networks $T^{A \to B}$ and $T^{B \to A}$. In parallel, two synthesis encoders $G(x^A)$ and $G(x^B)$ extract modality-invariant latent representations through the synthesis decoders $F^{A \to B}$ and $F^{B \to A}$, mapping the representations extracted by $G(x^A)$ and $G(x^B)$ to the synthesed images $\hat{x}^B$ and $\hat{x}^A$, respectively.

### 3.1 Architecture

Our model involves the following components (Figure 1): a) one synthesis encoder G, which extracts modality-invariant representations $G(x^A)$ and $G(x^B)$; b) two synthesis decoders $F^{A \to B}$ and $F^{B \to A}$, which map $G(x^A)$ and $G(x^B)$ to $\hat{x}^B = F^{A \to B}(G(x^A))$ and $\hat{x}^A = F^{B \to A}(G(x^B))$ (synthesised images); c) two transformation networks, $T^{A \to B}$ and $T^{B \to A}$, which model the transformation fields $\varphi^{A \to B} = T^{A \to B}(G(x^A), G(x^B))$ and $\varphi^{B \to A} = T^{B \to A}(G(x^B), G(x^A))$. Note that during training, $G(x^A)$ and $G(x^B)$ are warped into $G(x^A) \circ \varphi^{A \to B}$ and $G(x^B) \circ \varphi^{B \to A}$, before used to generate the synthesised images: $\hat{x}_T^B = F^{A \to B}(G(x^A) \circ \varphi^{A \to B})$ and $\hat{x}_T^A = F^{B \to A}(G(x^B) \circ \varphi^{B \to A})$.

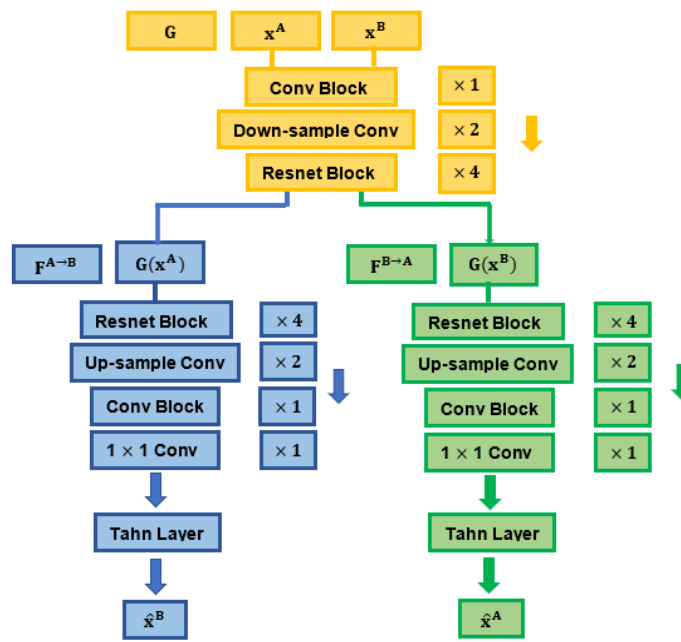

**Figure 2)** Details of the encoder and decoders used for the synthesis process.

**3.2 Encoder and Decoder for Synthesis**

The structures of the encoder and decoders were inspired by the design of CycleGAN [41], which includes a set of downsampling and upsampling convolutional layers connected through a series of Resnet blocks.

We modified the architecture of the CycleGAN generator, by adding two decoders in the generator and by eliminating the discriminator (instead of having one decoder in the generator followed by a discriminator, in the original CycleGAN, Figures 1 and 2) [41]. In our model, we did not need to recruit adversarial learning through the discriminator, as by involving two decoders and geometrical mapping via our STN-derived deformations in the cross-domain synthesis learning, our model was designed to match the (pixel intensity and geometrical) distribution of generated images to the data distribution in either target domain.

The encoder G contained one convolutional block (7 × 7, 64) with stride 1, two downsampling convolutional layers (3 × 3, 128; 3 × 3, 256) with stride 2, and four Resnet blocks of the same size (3 × 3, 256) with stride 1. The decoders contained four consecutive Resnet blocks of the same size (3 × 3, 256) with stride 1, two convolutional layers used for upsampling (3 × 3, 128; 3 × 3, 64) with stride 1 and a convolutional block (7 × 7, 3) with stride 1. The synthesised output was finally produced by a 1 × 1 convolutional layer, followed by a Tanh layer [22].

The downsampling layers (in the encoder) encode the images down to an abstract latent representation. In this process, the model learns weights that enable reducing the spatial resolution of the feature (image pixel) maps. The output feature maps derived from downsampling, are then passed through a series of Resnet blocks with skip connections, to further interpret and process their underlying information while addressing the vanishing gradient problem that may occur in large networks [20, 41]. Through the upsampling layers, the model learns to transpose (reverse) convolutions, decoding the latent representations back to the size of the ouput image.

## 3.3 Transformation Network

The transformation networks of our model $T^{A \to B}$ and $T^{B \to A}$ can learn both affine and non-rigid transformations (denoted as $\varphi_{af}$ and $\varphi_{nr}$, respectively, Figure 3). Each transformation network consists of a subnetwork for affine transformation followed by a subnetwork for nonrigid transformation, as shown in Figure 3.

To derive deeper geometrical inter-relationships from the processed data, we modified the start of the STN structure by adding three convolutional layers with instance normalization and standard LeakyReLU layers (Figure 3). In specific, the affine transformation network $T_{af}$ uses an initial convolutional layer of size (1 × 8, 7) and stride 2, followed by two "convolutional layer + ReLU + Instance Norm" blocks (the size of the first and second convolutional layers used were (8 × 10, 5; 16 × 20, 2) respectively, with stride 2). Following the two "convolutional layer + ReLU + Instance Norm" blocks, two Dense (fully-connected) layers were recruited to compute the parameters of the affine transformation matrix (2 × 3 for 2D; 3 × 4 for 3D transformations) [8]. Essentially, the input of the affine subnetwork is the concatenated features obtained from the two input images. The convolutional features derived from the third convolutional layer were resampled into a fixed size feature vector using Global Average Pooling [8, 22], followed by two fully connected layers which werer used to generate the transformation matrix.

The non-rigid transformation network $T_{nr}^{A \to B}$ receives the affinely transformed feature $G(x^A) \circ T_{nr}^{A \to B}$ and the untransformed feature $G(x^B)$ as inputs, and subsequently processes them in parallel layers (Figure 3). The features extracted are finally concatenated and used as inputs into a ResNet block with instance normalization and LeakyReLU layers. To produce the non-rigid deformation $\varphi_{nr}^{A \to B}$, a final convolutional layer and a Tanh activation function are implemented. The Tanh layer is performed on a normalized coordinate system: a coordinate $p \in [-1, 1]$ n exists for an n-D image [22].

As described, we factorised two separate transformation networks for each encode-decoder channel in our FIRE model which means that the above process described for the A→B (affine and non-rigid) transformations are identical for the B→A affine and non-rigid transformations as well.

## 3.4 Training Procedure and Overall Loss function

Our FIRE model is trained for simultaneously learning mutually inverse registration and synthesis tasks across both "A→B" and "B→A" directions.

The A→B image synthesis process synthesized images $\hat{x}^B$ and $\hat{x}^B_T$, so that $\hat{x}^B$ is registered with x$^A$ and $\hat{x}^B_T$ is identical to the target image x$^B$. The A→B transformation learned by T$^{A→B}$ is applied to the features G(x$^A$) for synthesis purposes, and subsequently to the image x$^A$ for registration.

The B→A image synthesis process performs backward registration and synthesis. All parameters of the entire network are updated end-to-end in the training process, and only G and a single transformation network are required to deform the moving image in the testing phase, for either modality input image.

To train the proposed FIRE model, a synthesis loss (Lsyn) and a registration loss (Lreg) are recruited. A regularisation process (R) is introduced, to perform topology-preserving deformation and spatial smoothing [22]. Combining these terms, the overall loss function of the proposed model is:

$$\mathcal{L} = \mathcal{L}syn + \mathcal{L}reg + \mathcal{R} \qquad (1)$$

As described in the next subsections, we formulate most of the terms in Lsyn, Lreg and R using root-mean-square (RMS) calculated on normalized inputs. As a result, apart from regularising the smoothness of the non-rigid deformation field, no hyperparameter is required to balance different losses, as detailed below.

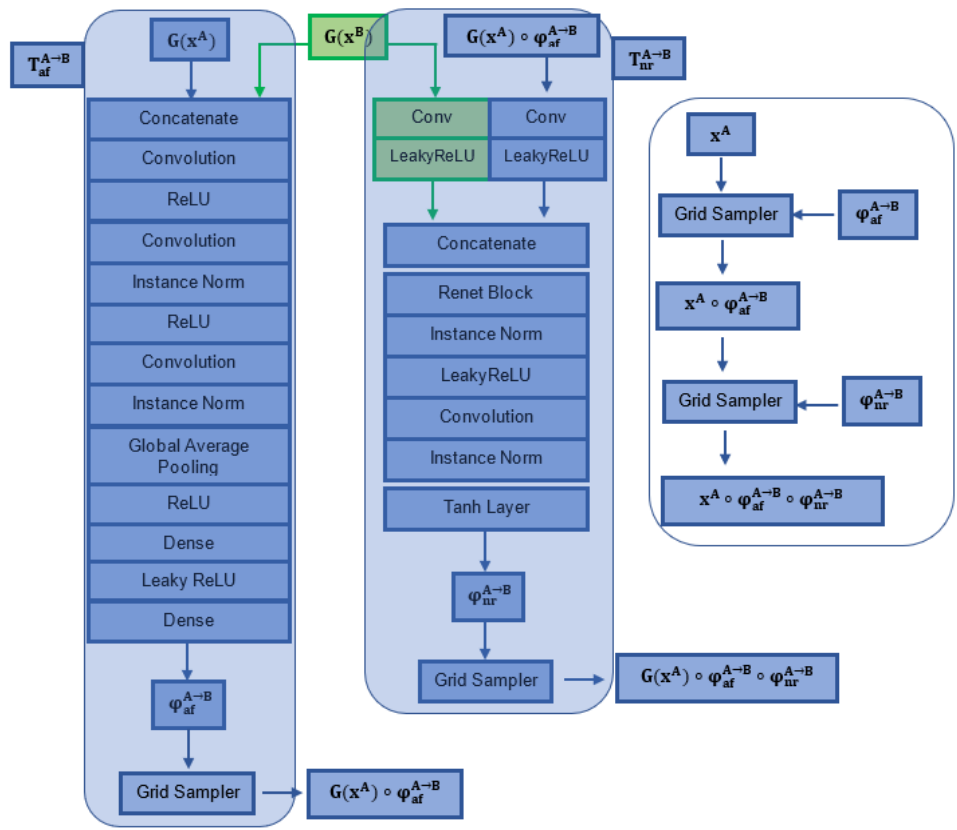

**Figure 3)** Affine and non-rigid spatial transformation networks incorporated in the FIRE model.

**3.5 Synthesis Loss**

There are four synthesis loss terms involved, each one supporting a different purpose (see Figure 4). First, to perform accurate cross-domain synthesis, the following synthesis accuracy loss was defined (by implementing the RMS error):

$$\mathcal{L}syn, acc = \text{RMS}(\hat{x}_T^B, x^B) + \text{RMS}(\hat{x}_T^A, x^A) \tag{2}$$

Note that for image synthesis, the synthesized images $\hat{x}_T^A$ and $\hat{x}_T^B$ aim to be identical to the target images $x^A$ and $x^B$ respectively, which is defined in equation 2.

Second, G is designed to learn modality-invariant features. To represent this, we defined the following feature loss, which aims to minimize the RMS error between $G(x^A)$ and $G(x^B) \circ \varphi^{B \rightarrow A}$, as well as $G(x^B)$ and $G(x^A) \circ \varphi^{A \rightarrow B}$:

$$\mathcal{L}syn, fea = \text{RMS}(G(x^A), G(x^B) \circ \varphi^{B \rightarrow A}) + \text{RMS}(G(x^B), G(x^A) \circ \varphi^{A \rightarrow B}) \quad (3)$$

Equation 3 was designed to minimise the error between the G outcomes from either modality, and the transformations learned in the G outcomes of the other modality.

Third, the cycle-consistency loss used in CycleGAN [20, 41] has been proved to be critical for its superior performance in cross-domain image synthesis. For robust cross-modality synthesis performance, a cycle-consistency loss was therefore designed:

$$\mathcal{L}syn, cyc = \text{RMS}(F^{B \rightarrow A}(G(\hat{x}^B)), x^A) + \text{RMS}(F^{A \rightarrow B}(G(\hat{x}^A)), x^B) \quad (4)$$

Essentially, equation 4 encourages $F^{B \rightarrow A}(G(\hat{x}^B)) = x^A$ and $F^{A \rightarrow B}(G(\hat{x}^A)) = x^B$, enforcing these mappings to be reverses of each other (by minimising the error of either mapping procedure to the inputs, $x^A$ and $x^B$).

Finally, aligning $x \cdot$ and $\hat{x} \cdot$ is important, to transfer the geometric correspondence from either $x^A$ or $x^B$ (learned through the transformation networks), to the synthesis process. To this end, the synthesis alignment loss was defined as:

$$\mathcal{L}syn, align = \text{RMS}(G(x^A), G(\hat{x}^B)) + \text{RMS}(G(x^B), G(\hat{x}^A)) \quad (5)$$

The entire FIRE synthesis loss was [22] (Figure 4):

$$\mathcal{L}syn = \mathcal{L}syn, acc + \mathcal{L}syn, fea + \mathcal{L}syn, cyc + \mathcal{L}syn, align \quad (6)$$

## 3.6 Registration Loss

To perform synthesis and registration, features extracted by G were transformed and registration was then performed through the following transformations to the input images $x^A$ and $x^B$: $\varphi^{A \to B} = \varphi_{af}^{A \to B} \circ \varphi_{nr}^{A \to B}$ and $\varphi^{B \to A} = \varphi_{af}^{B \to A} \circ \varphi_{nr}^{B \to A}$. Because the output of the synthesis process $F^{\cdot \to \cdot}(G(\cdot))$ is aligned with its input, the synthesized image obtained from a transformed $x^A$ image should be identical to $x^B$. Based on this, we defined the following registration accuracy loss:

$$\mathcal{L}reg, acc = \text{RMS}(F^{A \to B}(G(x^A) \circ \varphi^{A \to B})), x^B) + \text{RMS}(F^{B \to A}(G(x^B) \circ \varphi^{B \to A})), x^A) \tag{7}$$

Finally, for inverse-consistent registration, the transformations $\varphi^{A \to B}$ and $\varphi^{B \to A}$ should be mutually inverse thus, the following inverse-consistency loss was defined:

$$\mathcal{L}reg, ic = \text{RMS}(x^A, x^A \circ \varphi^{A \to B} \circ \varphi^{B \to A}) + \text{RMS}(x^B, x^B \circ \varphi^{B \to A} \circ \varphi^{A \to B}) \tag{8}$$

This inverse-consistent loss can be seen as a transformation-oriented cycle-consistency loss, which encourages the composition of mutual mappings from the moving image to the fixed image on a bi-directional mode. For example, for $x^A$ (moving image) to $x^B$ (fixed image) registration, it minimizes the error for both $\varphi^{A \to B}$ and $\varphi^{B \to A}$ mappings.

The overall entire registration loss was computed as:

$$\mathcal{L}reg = \mathcal{L}reg, acc + \mathcal{L}reg, ic \tag{9}$$

In practice, the non-rigid transformation field $\varphi_{nr}^{\cdot}$ was calculated on $G(x^{\cdot})$, and linearly resampled before applied to the image $x^{\cdot}$.

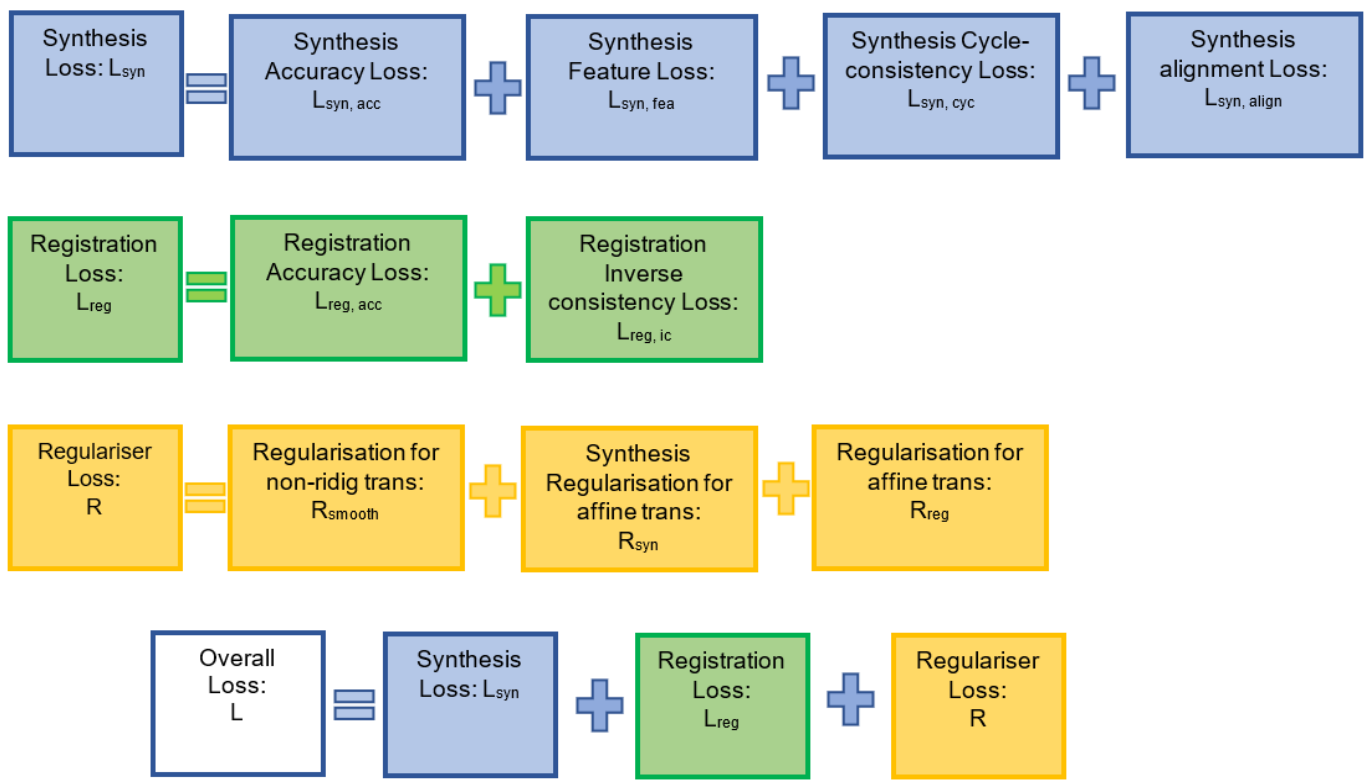

**Figure 4)** Graphical description of all losses defined in the FIRE model. Full descriptions and mathematical details are provided in the subsections 3.5-3.7 (in the Methods).

**3.7 Regularisation**

Inspired by conventional image registration in which regularisation is commonly an integral part [11, 22], we also add regularisation terms in both affine and non-rigid transformations. The main novelty in terms of regularisation in our work, is that we incorporate separate regularization terms for both affine and non-rigid transformations in our model. At that stage, it is important to note a fundamental difference between affine and non-rigid transformation [11, 30-33]: affine registration recruits global transformation fields, whilst non-rigid registration recruits mainly local transformation fields [11]. Non-rigid registration may contain affine transformations. However, affine registration should not involve non-rigid transformations.

In that context, we first add a bending energy penalty term to reinforce smooth displacements for both non-rigid transformation fields $\varphi_{nr}^{A \to B}$ and $\varphi_{nr}^{B \to A}$, as follows:

$$\mathcal{R}smooth = \| \nabla^2 \varphi_{nr}^{A \to B} \|^2 + \| \nabla^2 \varphi_{nr}^{B \to A} \|^2 \tag{10}$$

where $\nabla$ is the Laplacian operator.

Second, because our FIRE model is designed to also calculate affine registrations, we propose to induce a separate regularization term, to eliminate non-rigid transformations when affine tranformation is present in the data. During image synthesis, the affinely transformed features $G(x^A) \circ \varphi_{af}^{A \to B}$ and $G(x^B) \circ \varphi_{af}^{B \to A}$, can be used as inputs into the synthesis decoders to obtain $F^{A \to B}(G(x^A) \circ \varphi_{af}^{A \to B})$ and $F^{B \to A}(G(x^B) \circ \varphi_{af}^{B \to A})$. The regularisation of the synthesis is then computed as:

$$\mathcal{R}syn = \mathrm{RMS}(x^B, F^{A \to B}(G(x^A) \circ \varphi_{af}^{A \to B})) + \mathrm{RMS}(x^A, F^{B \to A}(G(x^B) \circ \varphi_{af}^{B \to A})) \tag{11}$$

Similarly, a regularisation for registration, $\mathcal{R}reg$, is defined using the affinely deformed $x^A$ and $x^B$:

$$\mathcal{R}reg = \mathrm{RMS}(x^B, F^{A \to B}(G(x^A \circ \varphi_{af}^{A \to B}))) + \mathrm{RMS}(x^A, F^{B \to A}(G(x^B \circ \varphi_{af}^{B \to A}))) \tag{12}$$

The overall regularisation of the FIRE model was [22]:

$$\mathcal{R} = \mathcal{R}syn + \mathcal{R}reg + \lambda \mathcal{R}smooth \tag{13}$$

where $\lambda$ is the scaling parameter for $\mathcal{R}smooth$, which is the only hyperparameter used in the FIRE loss. We should note that to register n-D images, $\lambda = 2^{2n}/10N$, where N is the number of points in the input image.

It is important to also emphasise that when non-rigid transformation is present, the model uses all 3 parameters $\mathcal{R}_{syn}, \mathcal{R}_{reg}$ and $\mathcal{R}_{smooth}$, from equation 13. When affine registration is present, the model focuses on using the $\mathcal{R}_{syn}$ and $\mathcal{R}_{reg}$ terms (since the Laplacian operator is not activated during affine registration).

**3.9 Optimization**

Computing Lreg requires to input the transformed images into G, which creates a circular computing graph. Furthermore, different networks in the proposed FIRE model show different behaviour in the training process. For example, $T_{af}^{\cdot \rightarrow \cdot}$ is more sensitive to changes in L, compared to G and $F^{\cdot \rightarrow \cdot}$.

To address this issue, we implemented three Adam optimizers to separately optimise parameters for a) $T_{af}^{\cdot \rightarrow \cdot}$, b) $T_{nr}^{\cdot \rightarrow \cdot}$ and c) the synthesis encoder/ two decoders. We used a uniform training procedure regardless of the size of the datasets. To optimize speed of convergence, learning rates for training $T_{af}^{\cdot \rightarrow \cdot}$ and $T_{nr}^{\cdot \rightarrow \cdot}$ and G/ $F^{\cdot \rightarrow \cdot}$ were set to $10^{-5}$, $5 \times 10^{-5}$ and $10^{-4}$, respectively. The FIRE model was trained end-to-end for 144,000 iterations for both datasets (empirical observation 1: for both brain and cardiac data, the total duration of the training phase was approximately 15-18 hours on a Tesla P40 GPU with 24G memory (about 7.5-9 hours per data set); empirical observation 2: for the cardiac data, model convergence was reached in the first 2-3 hours, but the model was kept under training for all 144,000 iterations).

## 4. Experiments

The performance of our proposed model was evaluated using multi (for inter-modality registration)- and single (for intra-modality registration)-modal MR data.

### 4.1 MRBrainS Data

For multi-modality registration, the training data from the the MRBrains13 (http://mrbrains13.isi.uu.nl/) and the MRBrains18 (https://mrbrains18.isi.uu.nl/) Challenges were fused. The fused dataset contained multi-modality brain MR data.

In detail, the dataset consisted of 3D $T_1$-weighted, $T_2$-Fluid-attenuated IR ($T_2$-FLAIR) and inversion recovery (IR) data from 12 subjects, acquired using 3T MRI. The 3D $T_1$, $T_2$-FLAIR and IR datasets included 192, 48 and 48 slices per patient, respectively. The 3D $T_1$ and IR data were already co-registered to the $T_2$-FLAIR data. Images across all modalities had a voxel size of 0.958 × 0.958 × 3.000 mm$^3$.

To assess model performance, manual annotations from three brain anatomical structures were used: the brain stem (BS), the cerebellum (Ce) and the white matter (WHM). For the training, validation and testing processes, we used all MRI slices from 8, 1 and 3 patients, respectively (Table 1). To perform 3D and 2D registration, all data were resampled to 1.28mm$^3$ per voxel.

In the registration process, 2D and 3D registration was performed between $T_1$ and $T_2$-FLAIR data, and 2D registration between IR and $T_2$-FLAIR data. During training, moderate-to-strong (20-50% change across at least one dimension) affine and non-rigid transformations were randomly applied to the moving and fixed images. In the testing phase, each of the $T_1$ and IR data were randomly transformed 20 times, and were subsequently allowed to be registered to the corresponding $T_2$-FLAIR data.

### 4.2 ACDC Data

To perform intra-modality registration, 4D cardiac cine MRI data from the 2017 ACDC (https://www.creatis.insa-lyon.fr/Challenge/acdc) Challenge were used. Note that the 4$^{th}$ dimension here describes temporal resolution. The voxel size was between 1.37-1.68 × 1.37-1.68 × 5-8 mm$^3$, and each 4D image has 28-40 phases covering the cardiac cycle.

In specific, the training dataset contained MRI data from 100 patients with different cardiovascular pathologies, with manual annotations of the myocardium and the left ventricle per patient at two cine-MR phases (images) per slice.

To train the FIRE model, we used all MRI phases per slice. To test the model, only the two annotated phases per slice were evaluated. For the training, validation and testing process, MRI data from 60, 10 and 30 patients were used, respectively (Table 1).

**4.3 Evaluation Metrics and Baselines**

Both the MRBrainS and ACDC datasets provided manual annotations (ground truths). To evaluate our method, we used the Dice metric to measure the overlap of the moving and fixed annotations. Increased Dice scores represent high registration performance, and vice versa.

Although there are recent developments in image registration (described in the Introduction and detailed in the Related work), the mutual information (MI) and Symmetric Normalization (SyN) techniques through using the Advanced Normalization Toolbox (ANTs) (http://stnava.github.io/ANTs/) [37], are still considered the current reference standard techniques for affine and non-rigid registration, respectively [4, 9-18, 30-41].

Hence, the proposed model is evaluated against the standard MI and SyN techniques, to assess affine and non-rigid registration performance in multi-slice multi-modality brain and multi-phase/multi-slice single-modality cardiac MRI data, respectively.

**Table 1.** Total number of images for training, validation and testing per brain and cardiac dataset. Note that for the cardiac cata, only 2 images per cardiac slice contained annotations and were evaluated at testing.

| Organ | Sequence | Training | Validation | Testing |
|---|---|---|---|---|
| **Brain** | $T_1$ | 1536 | 192 | 576 |
| | IR | 384 | 48 | 144 |
| | $T_2$-FLAIR | 384 | 48 | 144 |
| **Cardiac** | Cine | 14,400 | 240 | 480 |

## 5. Results

### 5.1 Inter-modality registration

Initially, 2D and 3D $T_1$ to $T_2$-FLAIR image registration was examined. The proposed model was consistent in achieving higher scores against the SyN method, for all brain anatomical areas investigated (Table 2). As presented in Table 2, the proposed model outperformed the SyN method across all 2D and 3D $T_1$ to $T_2$-FLAIR image registration experiments examined. These results were maintained when both affine and deformable image registration were evaluated.

A visual representation of the $T_1$ to $T_2$-FLAIR registration is illustrated in Figure 5. In this representation, the FIRE model shows better alignment between the outer contour (in the extracerebral space) outlining the cerebrospinal fluid (shown with blue) and the actual brain tissue delineation, versus the Syn technique.

Subsequently, IR to $T_2$-FLAIR registration was evaluated. Our proposed model consistently reached improved registration performance against the Syn method: a mean Dice score of 0.68 (0.3), 0.69 (0.2) and 0.70 (0.3) for the BS, Ce and WHM brain structures was reached, respectively. A visual representation of the IR to $T_2$-FLAIR registration is illustrated in Figure 6.

It is important to note that the average Dice score obtained across all brain anatomical areas was below 0.45, when the SyN technique was evaluated: there was a mean Dice score of 0.43 (0.2), 0.42 (0.3) and 0.44 (0.3) for the BS, Ce and WHM brain structures, respectively. These results were repeated when a grid-search for the Syn method within ANTs was carefully examined, making it impossible to derive a visual alignment of the IR to $T_2$-FLAIR registration.

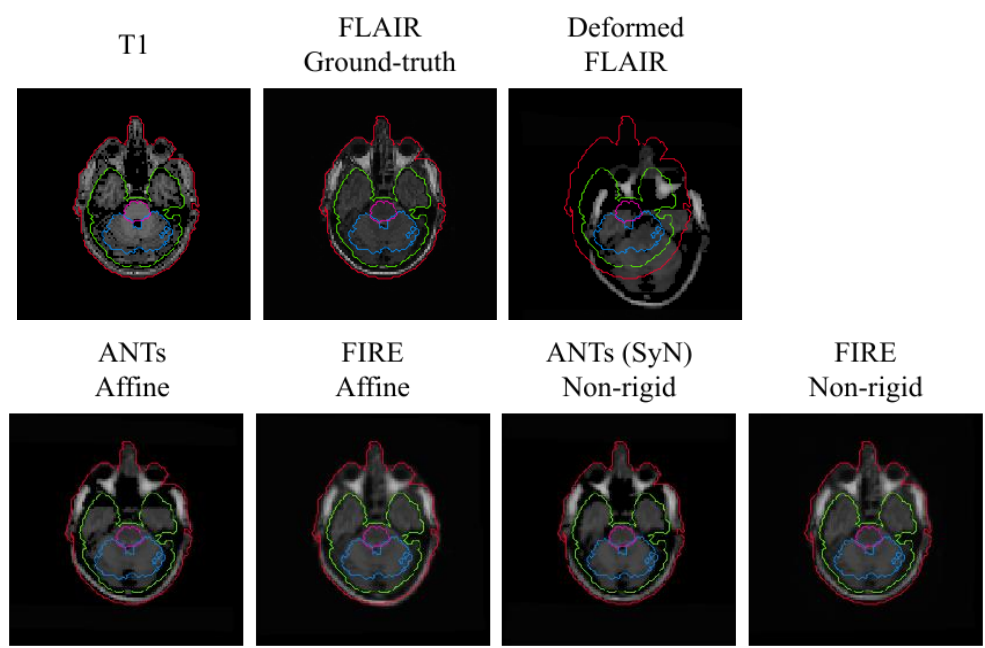

**Figure 5)** Visual representation of the $T_1$ to $T_2$-FLAIR registration using the MRBrainS data. Brain stem, cerebellum and white matter used to assess registration performance are illustrated with purple, blue and green, respectively.

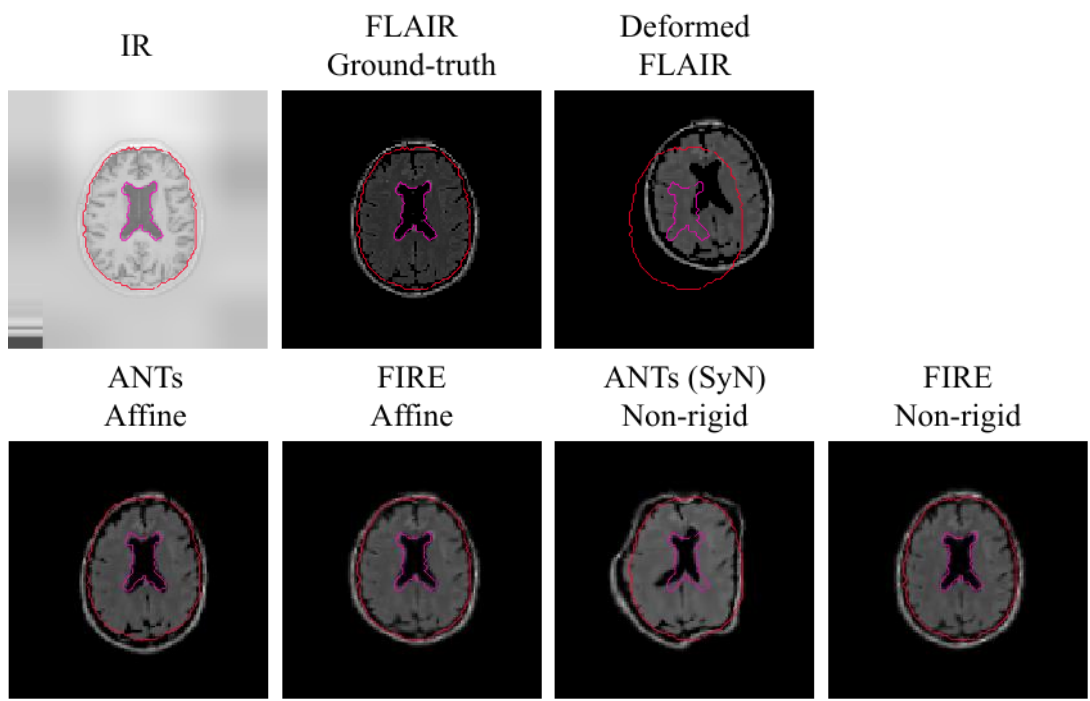

**Figure 6)** Visual representation of the IR to the $T_2$-FLAIR registration using the MRBrainS data. To increase clarity due to having less anatomical information in IR and $T_2$-FLAIR data, we use the outer brain region and ventricles (with red and purple respectively) to visually demonstrate registration results. Note that the upper left image shows an IR image before pixel normalization occurred during model fitting.

**Table 2.** Results obtained from the 2D and 3D $T_1$ to $T_2$-FLAIR registration, on the MRBrainS data. Dice scores were measured on the brain stem (BS), cerebellum (CE) and white matter (WHM). Standard deviations are included within the parenthesis.

| Data | Object | unaligned | ANTs-affine | FIRE-affine | ANTs-SyN | FIRE |
|---|---|---|---|---|---|---|
| 2D | BS | 11.62 (6.1) | 61.25 (3.7) | 62.90 (4.1) | 78.73 (7.3) | **80.68 (7.7)** |
| 2D | CE | 7.17 (4.4) | 63.32 (3.2) | 64.36 (4.0) | 75.72 (8.1) | **76.96 (7.3)** |
| 2D | WHM | 14.29 (7.5) | 59.12 (4.5) | 59.97 (4.4) | 81.36 (6.0) | **84.18 (3.7)** |
| 3D | BS | 27.15 (9.2) | 67.15 (3.1) | 69.81 (4.1) | 79.77 (6.7) | **81.08 (7.0)** |
| 3D | CE | 28.38 (9.5) | 68.38 (3.6) | 70.62 (3.7) | 86.00 (6.9) | **86.13 (7.2)** |
| 3D | WHM | 20.27 (9.3) | 60.27 (3.8) | 60.61 (3.8) | 72.33 (7.4) | **72.56 (7.1)** |

## 5.2 Intra-modality registration

Intra-modality registration was then investigated using 4D cardiac cine-MRI data. On the left ventricle anatomical areas, the FIRE model and the Syn method showed high and comparable Dice scores (Table 3, Figure 7). On the myocardial anatomical areas, the FIRE model marginally outperformed the Syn method (Table 3).

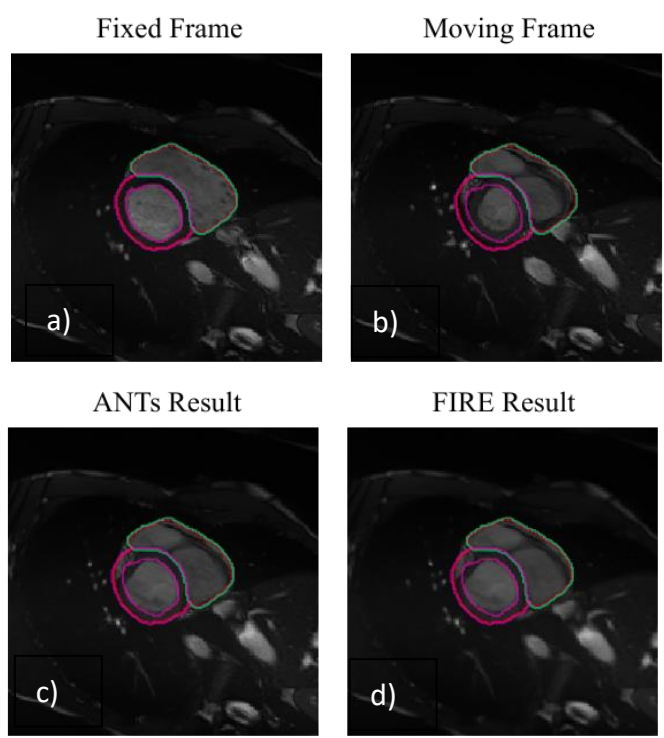

**Figure 7**) Illustration of the intra-modality image registration, using the ACDC data (the blue contour demonstrates the left ventricle delineation). Full diastole (a, c) and systole (b, d) are illustrated.

**Table 3.** Results on ACDC data. Dice scores computed on left ventricular endocardium (LVe) and myocardium (Myo). Standard deviations are included within the parenthesis. LVe: left ventricle, Myo: myocardial tissue.

| Object | unaligned | ANTs-SyN | FIRE |
|--------|-----------|----------|------|
| LVe | 65.75 (16.2) | **90.81 (4.3)** | 90.28 (5.5) |
| Myo | 51.95 (14.5) | 70.71 (5.6) | **71.86 (6.3)** |

### 5.3 Computational times

We methodologically assessed the computational costs required for each image registration process using the FIRE and the baseline SyN method implemented in ANTs. For both the FIRE model and SyN methods, the running times at the testing phase were calculated on a CPU system (Intel(R) Xeon(R) Silver 4112 CPU @ 2.60GHz, RAM: 256 GB). Furthermore, the GPU-accelerated performance of our FIRE model was also measured using a Tesla P40 GPU with 24G memory. Note that GPU acceleration is not available for the SyN method within ANTs.

Average running times for registering 3D and 2D data were obtained from 30 volumes, each volume containing 48 2D slices. For the 4D cardiac MR data provided in the ACDC dataset, we computed the average time for registering 10 3D volumes (of 10 slices each) at full systole to their first cardiac cine MR phase representing full diastole, therefore using 100 image pairs in total. All results are shown in Table 4.

Under the CPU mode, our FIRE model is consistently faster compared to the baseline Syn method, for both affine and non-rigid registration of the 4D cardiac data. The FIRE model shows comparable speed with the baseline in non-rigid registration of 3D and 2D brain data, whilst it was only slower for affine registration of the 3D brain data (Table 4).

When accelerated by a GPU system, our FIRE model reduced considerably the computational cost (for at least 30 times), when examined on 2D and 3D brain data. The FIRE model saved over 99.7% running time compared to the SyN method, while achieving higher accuracy in non-rigid registration of the 4D ACDC data.

**Table 4.** Mean computational times of the FIRE and the Syn method in the testing phase. The parenthesis show standard deviations.

| Data | Registration | Multi-modal | Number of Images used | FIRE-GPU (sec) | FIRE-CPU (sec) | SyN (sec) |
|---|---|---|---|---|---|---|
| MRBrainS (2D) | Affine | Yes | 1,440 | 0.305 (0.01) | 1.1856 (0.03) | **0.3375 (0.06)** |
| | Non-rigid | Yes | | 0.352 (0.01) | 1.5370 (0.04) | **1.3987 (0.09)** |
| MRBrainS (3D) | Affine | Yes | | 0.454 (0.06) | 37.939 (0.48) | **4.3542 (0.13)** |
| | Non-rigid | Yes | | 0.555 (0.07) | **29.004 (0.34)** | 30.069 (0.49) |
| ACDC (4D) | Affine | No | 200 | 4.873 (0.08) | **199.54 (1.26)** | 224.78 (1.89) |
| | Non-rigid | No | | 5.026 (0.09) | **200.16 (1.78)** | 1887.30 (2.25) |

## 6. Discussion

We demonstrate that the FIRE method is an efficient DL model that performs accurate and fast inter- and intra-modality affine and is capable for both affine and non-rigid image registration. The FIRE model efficiently learns inverse-consistent topology-preserving deformations when evaluated on 2D and 3D multi-modal brain MR and 4D single-modal cardiac MR data, showing higher accuracy and robustness against the reference standard Syn method.

### 6.1 Inter-modality registration on brain MRI

We showed that when 2D as well as 3D image registration were examined on $T_1$ to $T_2$-FLAIR registration, the proposed model was consistent in showing higher performance against the Syn technique, across all brain structures assessed (Table 2). This finding indicates that the proposed model can robustly model different anatomical and semantic features in these multi-tissue/multi-modal MR data thus, behaving as a brain tissue- and MR modality-agnostic model (for $T_1$ to $T_2$-FLAIR registration).

The IR to $T_2$-FLAIR image registration was compromised by the lack of anatomical information in either input modality. Implementing the reference standard Syn technique, it was not possible to reach accurate image registration results on quantitative assessments (all DICE scores were lower than 0.45), whilst we did not reach optimal alignments on visual assessments (Figure 6). It is important to note that the Syn method is the reference standard baseline technique for model evaluation in image registration [3, 4, 9-18, 30-41]. Despite this, out FIRE model outperformed the Syn method and was able to achieve moderate mean Dice scores across all brain structures (subsection 5.1).

Moreover, when examined in the testing phase, the FIRE model showed comparable speed with the baseline in non-rigid registration of 3D and 2D brain data (Table 4). The only case for which the FIRE model was slower, was when affine registration of 3D brain data was measured. Nevertheless, all FIRE model-derived affine and non-rigid registration processes for all 3D brain volumes were performed in less than 40 sec.

The FIRE model can therefore perform multi-modal affine and non-rigid registration on a memory-saving mode, without requiring supercomputers or GPU systems in the testing phase.

## 6.2 Intra-modality registration on cardiac MRI

We demonstrated that the FIRE model can optimally model deformable transformations in cardiac cine-MR data (Table 3). Although there were only 2 images segmented across each cine-MR dataset (used in the testing phase), these represent the maximum geometric difference within each cardiac data-set (one segmented image is always on full diastole and the other on full systole within ACDC data). Moreover, despite the local displacements between diastole and systole were relatively small (in the 3D space), the tissue deformation is the largest within cardiac cine-MR data. In other words, this means that the cardiac tissue is non-uniformly non-rigidly deformed across phases, with the maximum deformation occurring between diastole and systole (Figure 7). Hence, while using all cardiac phases (in between diastole and systole) in the training phase (representing 4D information), our results in the testing phase show that the FIRE model can optimally learn to perform accurate registration between largely deformed geometries (different cardiac shapes between systole and diastole).

In addition, the computational cost for the FIRE model was substantially lower compared to the Syn method, when non-rigid 4D registration of cardiac MR data was examined. Thus, the FIRE model demonstrated overall improved registration speed when dealing with 4D data, which can be particularly important across numerous cardiac MR applications involving time series [41-44]. To our knowledge, the only other technique that can effectively perform inverse consistent registration directly in the training phase, is the CycleMorph model [33]. In our work, we define the output deformation fields with simple inverse-consistency constraints, devising a model that can inherently learn topology-preserving image registration in a consistent mode (via the synthesis encoder G and a single transformation network).

## 6.3 FIRE model explainability

Despite recent developments in DL, it is known that inter-modality image registration is not yet widely investigated using deep networks [30, 31, 33, 34]. A major reason is the absence of explainability regarding why a DL model learns and/or where fails to learn anatomical image information and semantics. This becomes more evident in image registration, as the lack of standardized annotations (ground truths) may have diversified methods for model evaluation and thus, discouraged model explainability [5-7, 26-29, 30-34, 35, 38, 41]. Our main consideration here is that DL explainability can help to interpret model-data interrelationships before (or beyond) deriving class activation maps [45, 46], which can encourage thorough (and explainable) DL investigations in inter-modality image registration. We showed that the FIRE model can learn anatomical and semantic representations across modalities, and demonstrated improved performance for inter-modality image registration, versus the SyN method.

It is important to note that $T_1$, $T_2$-FLAIR and IR are routinely used MR sequences, but there are fundamental disparities in terms of the MR physics involved and thus, their imaging content [47]. These MR modalities are optimised to provide different imaging information in clinical MRI: $T_1$ is designed to provide detailed anatomical information, whereas IR and $T_2$-FLAIR are primarily implemented to extract functional information [47]. In $T_1$ to $T_2$-FLAIR registration, our cross-domain synthesis component (encoder/decoders) can effectively leverage anatomical information from the $T_1$ data, and can subsequently map this information to complementary functional $T_2$-FLAIR information. This complementary anatomical-functional information is compromised in the IR to $T_2$-FLAIR data registration, during cross-domain synthesis. Hence, not the anatomical information per se, but the presence of (heterogeneous) complementary anatomical-functional information makes the multi-modal transformations invertable in relation to each other, which in turn encourages our model to produce inverse-consistent topology-preserving image registrations. Although not in the scope of the current study, further work is underway to quantify this complementary information in our model, through gradient-weighted class activation mapping (Grad-CAM) [48], across each of the model main components (enconder, STN, decoders).

We should also highlight that there can be multiple methodologies by which it can become possible to enrich anatomical information within a dataset when missing, to enhance inverse-consistency through complementary information. For example, an accurate semi-supervised inter-modality learning model could be explored to derive anatomical annotations and guide the registration process [49, 50] hence, potentially adapting our technique for functional (non-anatomical) MR modalities [49, 50]. Other active contour-based segmentation algorithms for fast image segmentation have shown excellent segmentation accuracies even for images with large noise interference and intensity inhomogeneities [51-53], and therefore may be applicable to enrich anatomical information in multi-modal registration of imaging modalities with decreased anatomical information.

## 7. Conclusions

We have demonstrated a robust method that efficiently models inter- and intra-modality image registration, through bi-directional cross-domain synthesis and factorised spatial transformation. To our knowledge, parallelizing cross-domain synthesis while modelling spatial transformation to dynamically learn anatomical and latent representations simultaneously across modalities, has not been previously investigated.

Our work showed the efficiency of our methodology as we improve intra- and inter-modality registration by maintaining fast computational times in the testing phase, across all registration tasks.

In broader terms, the main significance of our work is the simultaneous bi-directional cross-domain image synthesis and spatial transformation per synthesis channel, which enables to diversify our technique across numerous multi-modality medical image analysis scenarios. Through providing a model explainability framework, we can suggest that it is possible to adapt our technique to learn anatomical-semantic information when we have at least one dataset with enriched anatomical information. This means that we can customize our methodology to address image registration on additional anatomically-enriched imaging data, such as with computed tomography (CT) or ultrasound data, next to MRI.

To date, the main limitation of our work is that we have only assessed multi-modal MRI data. However, we examined two organ areas (brain and heart), intra- and inter-modality registration, and investigated 2D, 3D and 4D registrations frames. This has inspired our future work as we currently investigate model generalisation frameworks, to allow further multi-organ and multi-modal applications. In conclusion, here we show that our methodology demonstrated improved performance and efficiency against the current standard Syn method thus, presenting a versatile image registration technique that may have merit in the clinical setting.

**Acknowledgements**

The authors express their gratidute to Professor Sotirios Tsaftaris and Dr Agisilaos Chartsias for the thorough discussions at the initial stages of this work.

**Funding**

This study was supported in part by the British Heart Foundation (TG/18/5/34111, PG/16/78/32402), the European Research Council Innovative Medicines Initiative (101005122), the European Commission H2020 (952172), the Medical Research Council (MC/PC/21013), and the UKRI Future Leaders Fellowship (MR/V023799/1).

**Author contributions**

CW conceptualized the work. CW, GY and GP worked on data analysis, curation, engineering and investigation. Writing, review and editing were performed by all authors. GP edited the final draft and performed project management. All authors have read and approved submission of the manuscript.

**Data availability statement**

For our data analysis, we used publicly available datasets and we explicitly describe how to access these data. Code will be made available at GitHub.

**Conflicts of Interest**

GP is currently an employee of Pfizer. The other authors have no conflict of interest to disclose.

**References**

1. Rueckert D, Schnabel JA: Medical image registration. In: Biomedical image processing. Springer. 2010; 131–154.

2. Zhou SK, Greenspan H, Davatzikos C, et al. A review of deep learning in medical imaging: imaging traits, technology trends, case studies with progress highlights and future promises. Proceedings of the IEEE. 2021; 109(5): 820-838.

3. Fu Y, Lei Y, Wang T, Curran WJ, Liu T, Yang X. Deep learning in medical image registration: A review. Phys Med Biol. 2020; 65(20): 20TR01.

4. Haskins G, Kruger U, Yan P. Deep learning in medical image registration: A survey. Mach Vis Appl. 2020; 31: 8.

5. Cao X, Yang J, Zhang J, Nie D, Kim M, Wang Q, Shen D: Deformable image registration based on similarity-steered cnn regression. In: International Conference on Medical Image Computing and Computer-Assisted Intervention, Springer. 2017: 300-308.

6. Krebs J, Mansi T, Delingette H, Zhang L, Ghesu FC, Miao S, Maier AK, Ayache N, Liao R, Kamen A. Robust non-rigid registration through agent based action learning. In: International Conference on Medical Image Computing and Computer-Assisted Intervention, Springer. 2017: 344-352.

7. Roh´e MM, Datar M, Heimann T, Sermesant M, Pennec X. Svf-net: learning deformable image registration using shape matching. In: International Conference on Medical Image Computing and Computer-Assisted Intervention, Springer. 2017; 266–274.

8. Jaderberg M, Simonyan K, Zisserman A, et al. Spatial transformer networks. In: Advances in neural information processing systems. 2015; 2017-2025.

9. Fan JF, Cao XH, Wang Q, Yap PT, Shen DG. Adversarial learning for mono- or multi-modal registration. Med. Image Anal. 58: 101545.

10. Fan JF, Cao XH, Yap EA, Shen DG. BIRNet: brain image registration using dual-supervised fully convolutional networks Med. Image Anal. 54: 193–206.

11. de Vos BD, Berendsen FF, Viergever MA, Sokooti H, Staring M, Isgum I. A deep learning framework for unsupervised affine and deformable image registration. Medical image analysis. 2019; 52: 128-143.

12. Elmahdy MS, et al. Robust contour propagation using DL and image registration for online adaptive proton therapy of prostate cancer. Med. Phys. 2019; 46: 3329-43.

13. Lei Y, Fu Y, Wang T, Liu Y, Patel P, Curran WJ, Liu T, Yang X. 4D-CT deformable image registration using multiscale unsupervised DL. 2020; Phys. Med. Biol. 65: 085003.

14. Vos B, Berendsen F F, Viergever M A, Staring M and Išgum I 2017 End-to-end unsupervised deformable image registration with a convolutional neural network. Deep Learning in Medical Image Analysis and Multimodal Learning for Clinical Decision Support. DLMIA 2017, ML-CDS. Lecture Notes in Computer Science, ed M Cardoso (Berlin: Springer). 2017; 10553. 204–12.

15. Ferrante E, Oktay O, Glocker B, Milone DH. On the adaptability of unsupervised CNN-based deformable image registration to unseen image domains. Machine Learning in Medical Imaging (MLMI). Lecture Notes in Computer Science vol 11046, ed Y Shi, H I Suk and M Liu (Berlin: Springer). 2018; 294–302.

16. Sheikhjafari A, Noga M, Punithakumar K, Ray N. Unsupervised deformable image registration with fully connected generative neural network. Submission to 1st Conf. on Medical Imaging with Deep Learning (MIDL 2018). 2018.

17. Jiang Z, Yin F F, Ge Y, Ren L. A multi-scale framework with unsupervised joint training of convolutional neural networks for pulmonary deformable image registration. Phys. Med. Biol. 2019; 65: 015011.

18. Nguyen-Duc T, Yoo I, Thomas L, Kuan A, Lee WC, Jeong WK. Weakly supervised learning in deformable EM image registration using slice interpolation. IEEE 16th Int. Symp. on Biomedical Imaging (ISBI 2019). 2019; 670-3.

19. Christensen GE, Johnson HJ. Consistent image registration. IEEE Trans Med Imaging. 2001: 20(7): 568-582.

20. Qin C, Shi B, Liao R, Mansi T, Rueckert D, Kamen A. Unsupervised deformable registration for multi-modal images via disentangled representations. in Proc Inf Process Med Imag (IPMI). Springer. 2019; 249-261.

21. Zhang J. Inverse-consistent deep networks for unsupervised deformable image registration. arXiv preprint arXiv:1809.03443 2018.

22. Wang C, Guang Y, Papanastasiou G. FIRE: unsupervised bi-directional inter- and intra- modality registration using deep networks. International Symposium on Computer-Based Medical Systems (CBMS). IEEE. 2021.

23. Wu G, Kim M, Wang Q, Gao Y, Liao S, Shen D. Unsupervised deep feature learning for deformable registration of mr brain images. In: International Conference on Medical Image Computing and Computer-Assisted Intervention (MICCAI). Springer. 2013; 649-656.

24. Blendowski M, Heinrich MP. Combining mrf-based deformable registration and deep binary 3d-cnn descriptors for large lung motion estimation in COPD patients. *Int J* Comput Assist Radiol Surg. 2019; 14(1): 43-52.

25. Simonovsky M, Gutierrez-Becker B, Mateus D, Navab N, Komodakis N. A deep metric for multimodal registration. In: International conference on medical image computing and computer-assisted intervention (MICCAI), Springer. 2016; 10-18.

26. Liao R, Miao S, de Tournemire P, Grbic S, Kamen A, Mansi T, Comaniciu D. An artificial agent for robust image registration. In: Thirty-First AAAI Conference on Artificial Intelligence. 2017.

27. Ma K, Wang J, Singh V, Tamersoy B, Chang YJ, Wimmer A, Chen T. Multimodal image registration with deep context reinforcement learning. In: International Conference on Medical Image Computing and Computer-Assisted Intervention (MICCAI), Springer. 2017; 240-248.

28. Hu J, et al. End-to-end multimodal image registration via reinforcement learning. Med Image Anal. 2020; **68:**101878.

29. Cao X, Yang J, Wang L, Xue Z, Wang Q, Shen D. Deep learning based inter-modality image registration supervised by intra-modality similarity. In: International Workshop on Machine Learning in Medical Imaging, Springer. 2018; 55-63.

30. Balakrishnan G, Zhao A, Sabuncu MR, Guttag J, Dalca AV. An unsupervised learning model for deformable medical image registration. Proceedings of the IEEE Conference on Computer Vision and Pattern Recognition. 2018; 9252-9260

31. Balakrishnan G, Zhao A, Sabuncu MR, Guttag J, Dalca AV. Voxelmorph: a learning framework for deformable medical image registration. IEEE Trans Med Imaging. 2019; 38(8): 1788-1800.

32. Krebs J, Mansi T, Mailhé B, Ayache N, Delingette H. Learning structured deformations using diffeomorphic registration. arXiv preprint. arXiv:1804.07172. 2018.

33. Kim B, Kim DH, Park SH, Kim J, Lee JG, Ye JC. CycleMorph: Cycle consistent unsupervised deformable image registration. Med Image Anal. 2021; 71: 102036.

34. Dalca AV, Balakrishnan G, Guttag J, Sabuncu MR. Unsupervised learning for fast probabilistic diffeomorphic registration. International Conference on Medical Image Computing and Computer-Assisted Intervention (MICCAI), Springer. 2018; 729-738.

35. Dalca AV, Guttag J, Sabuncu MR. Anatomical priors in convolutional networks for unsupervised biomedical segmentation. In: Proceedings of the IEEE Conference on Computer Vision and Pattern Recognition. 2018; 9290–9299.

36. Sokooti H, de Vos B, Berendsen F, Lelieveldt BP, Isgum I, Staring M. Nonrigid image registration using multi-scale 3d convolutional neural networks. In: International Conference on Medical Image Computing and Computer-Assisted Intervention (MICCAI), Springer. 2017; 232–239.

37. Avants BB, Epstein CL, Grossman M, Gee JC. Symmetric diffeomorphic image registration with cross-correlation: evaluating automated labeling of elderly and neurodegenerative brain. Medical image analysis. 2008; 12(1): 26-41.

38. Yang X, Kwitt R, Styner M, Niethammer M. Quicksilver: Fast predictive image registration-a deep learning approach. NeuroImage. 2017; 158: 378-396.

39. Nielsen RK, Darkner S, Feragen A. Topaware: Topology-aware registration. In: International Conference on Medical Image Computing and Computer-Assisted Intervention (MICCAI), Springer. 2019; 364–372.

40. Zhu Y, Park T, Isola P, Efros AA. Unpaired image-to-image translation using cycle-consistent adversarial networks. In: Proceedings of the IEEE International Conference on Computer Vision. 2017; 2223–2232.

41. Wei D, Ahmad S, Huo J, Peng W, Ge Y, Xue Z, Yap PT, Li W, Shen D, Wang Q. Synthesis and inpainting-based MR-CT registration for image-guided thermal ablation of liver tumors. In: International Conference on Medical Image Computing and Computer-Assisted Intervention (MICCAI), Springer. 2019; 512–52.

42. Papanastasiou G, Williams MC, Kershaw LE, *et al.* Measurement of myocardial blood flow by cardiovascular magnetic resonance perfusion: comparison of distributed parameter and Fermi models with single and dual bolus. J Cardiovasc Magn Reson. 2015; 17, 17.

43. Papanastasiou G, Williams MC, Dweck MR. *et al.* Quantitative assessment of myocardial blood flow in coronary artery disease by cardiovascular magnetic resonance: comparison of Fermi and distributed parameter modeling against invasive methods. J Cardiovasc Magn Reson*.* 2016; 18, 57.

44. Papanastasiou G, Williams MC, Dweck MR, et al. Multimodality quantitative assessments of myocardial perfusion using dynamic contrast enhanced magnetic resonance and $^{15}$O-labelled water positron emission tomography imaging. IEEE Trans Radiat Plasma Med Sci. 2018; 2(3):259-271.

45. Linardatos P, Papastefanopoulos V, Kotsiantis S. Explainable AI: A Review of Machine Learning Interpretability Methods. *Entropy*. 2021; 23(1):18.

46. Zhou B, Khosla A, Lapedriza A, Oliva A, Torralba A. Learning deep features for discriminative localization. In Proceedings of the IEEE Conference on Computer Vision and Pattern Recognition, Las Vegas, NV, USA, 27–30 June 2016; pp. 2921–2929.

47. Biglands JD, Radjenovic A, Ridgway JP. Cardiovascular magnetic resonance physics for clinicians: part II. J Cardiovasc Magn Reson. 2012: 14, 66.

48. Selvaraju RR, Cogswell M, Das A, Vedantam R, Parikh D, Batra D. GradCAM: visual explanations from deep networks via gradient-based localization. In: Proceedings of the IEEE International Conference on Computer Vision 2017, pp. 618– 626.

49. Chartsias A, Joyce T, Papanastasiou G, Semple S, Williams M, Newby DE, Dharmakumar R, Tsaftaris SA. Disentangled representation learning in cardiac image analysis. Medical Image Analysis. 2019; 58:101535.

50. Chartsias A, Papanastasiou G, Wang C, Semple S, Newby DE, Dharmakumar R, Tsaftaris SA. Disentangle, align and fuse for multimodal and semi-supervised image segmentation. IEEE Trans. Med. Imaging. 2020; 40: 781–792.

51. Ding K, Xiao L, Weng G. Active contours driven by region-scalable fitting and optimized Laplacian of Gaussian energy for image segmentation. Signal Processing. 2017; 134:224-233.

52. Jin R, Weng G. Active contours driven by adaptive functions and fuzzy c-means energy for fast image segmentation. Signal Processing. 2019; 163: 1-10.

53. Weng G, Dong B, Lei Y. A level set method based on additive bias correction for image segmentation. Expert Systems with Applications. 2021; 185: 115633.